\documentstyle[aps]{revtex}
\input epsf
\def\jepsfbox#1{\typeout{#1} \epsfbox{#1}}
\def\plotone#1{\begin{center} \leavevmode
\epsfxsize=0.7\columnwidth \jepsfbox{#1} \end{center}}
\def\jcite#1#2{#1 \cite{#2}}
\def\tilde{\mathaccent"365}			
\def\ie{{\it i.e.~}}

\def\etal{{\it et al.~}}
\def\rmmat#1{{\hbox{\rm #1}}}

\newcommand{\be}{\begin{equation}}
\newcommand{\ee}{\end{equation}}
\newcommand{\ba}{\begin{eqnarray}}
\newcommand{\ea}{\end{eqnarray}}
\def\p{\partial}
\def\d{{\rm d}}

\def\pp#1#2{\frac{\p #1}{\p #2}}
\def\IM{{\rm Im}}   
\def\RE{{\rm Re}}   
\def\figref#1{Fig.~\ref{fig:#1}}
\def\eqref#1{Eq.~\ref{eq:#1}}
\begin{document}
\draft
\newcommand{\bfi}{{\bf B}}
\newcommand{\efi}{{\bf E}}
\newcommand{\dfi}{{\bf D}}
\newcommand{\hfi}{{\bf H}}
\newcommand{\lag}{{\cal L}}
\newcommand{\dLIII}{{\frac{\partial^3 \lag}{\partial I^3}}}
\newcommand{\dLII}{{\frac{\partial^2 \lag}{\partial I^2}}}
\newcommand{\dLI}{{\frac{\partial \lag}{\partial I}}}
\newcommand{\dLKKK}{{\frac{\partial^3 \lag}{\partial K^3}}}
\newcommand{\dLKK}{{\frac{\partial^2 \lag}{\partial K^2}}}
\newcommand{\dLK}{{\frac{\partial \lag}{\partial K}}}
\newcommand{\dLIK}{{\frac{\partial^2 \lag}{\partial I \partial K}}}
\title{Birefringence and Dichroism of the QED Vacuum}

\author{Jeremy S. Heyl \and Lars Hernquist\thanks{Presidential Faculty Fellow}}
\address{Lick Observatory, 
University of California, Santa Cruz, California 95064, USA}
\maketitle
\begin{abstract}
We use an analytic form for the Heisenberg-Euler Lagrangian
to calculate the birefringent and dichroic
 properties of the vacuum for arbitrarily strong wrenchless fields.
\end{abstract}

\section{Introduction}

In the presence of a strong external field, the vacuum reacts,
becoming magnetized and polarized.  The index of refraction, magnetic
permeability, and dielectric constant of the vacuum are
straightforward to calculate using quantum electrodynamic one-loop
corrections \cite{Klei64a,Klei64b,Erbe66,Adle71,Bere82,Miel88}.  In
this paper, we calculate the magnetic permeability and dielectric tensors of
an external electric or magnetic field of arbitrary strength in terms
of special functions.  We combine these general results to calculate
the complex-valued index of refraction as a function of field
strength.

\section{The Permeability and Polarizability of the Vacuum}

When one-loop corrections are included in the Lagrangian of the
electromagnetic field one obtains a non-linear correction term:
\be
\lag = \lag_0 + \lag_1.
\label{eq:lagdef}
\ee
Both terms of the Lagrangian can be written in terms of the Lorentz
invariants,
\be
I = F_{\mu\nu} F^{\mu\nu} = 2 \left ( |\bfi|^2 - |\efi|^2 \right )
\label{eq:Idef}
\ee
and
\be
K = \left (\frac{1}{2}
\epsilon^{\lambda\rho\mu\nu} F_{\lambda\rho} F_{\mu\nu} \right )^2 =
	- (4 \efi \cdot \bfi )^2,
\label{eq:Kdef}
\ee
following Heisenberg and Euler \cite{Heis36}.  We use Greek indices to
count over space and time components ($0,1,2,3$) and Roman indices to
count over spatial components only ($1,2,3$), and repeated indices
imply summation.

\jcite{Heisenberg and Euler}{Heis36} and \jcite{Weisskopf}{Weis36}
independently derived the effective Lagrangian of the
electromagnetic field using electron-hole theory.
\jcite{Schwinger}{Schw51} later rederived the same result using
quantum electrodynamics.  In Heaviside-Lorentz units,
Lagrangian is given by
\ba
\lag_0 & = & -{1 \over 4} I, \label{eq:lag0def} \\
\lag_1 & = & {e^2 \over h c} \int_0^\infty e^{-\zeta} 
{\d \zeta \over \zeta^3} \left \{ i \zeta^2 {\sqrt{-K} \over 4} \times
\phantom{ 
\cos \left ( {\zeta \over B_k} \sqrt{-{I\over 2} + i \sqrt{K}} \right ) 
\over
\cos \left ( {\zeta \over B_k} \sqrt{-{I\over 2} + i \sqrt{K}} \right ) } 
\right . \\*
\nonumber
& & ~~ \left . 
{ \cos \left ( {\zeta \over B_k} \sqrt{-{I\over 2} + i {\sqrt{-K}\over 2}} \right ) +
\cos \left ( {\zeta \over B_k} \sqrt{-{I\over 2} - i {\sqrt{-K}\over 2}} \right ) \over
\cos \left ( {\zeta \over B_k} \sqrt{-{I\over 2} + i {\sqrt{-K}\over 2}} \right ) -
\cos \left ( {\zeta \over B_k} \sqrt{-{I\over 2} - i {\sqrt{-K}\over 2}} \right ) } 
 + |B_k|^2 + {\zeta^2 \over 6} I \right \}.
\label{eq:lag1def}
\ea
where $B_k = E_k = {m^2 c^3 \over e \hbar} \approx 1.3 \times 10^{16}
\rmmat{\,V\,cm}^{-1} \approx 4.4 \times 10^{13}$\,G.

In the weak field limit Heisenberg and Euler \cite{Heis36} give
\be
\lag \approx -{1 \over 4} I + E_k^2 {e^2 \over h c} \left [
{1 \over E_k^4} \left ( {1 \over 180} I^2 - {7 \over 720} K \right
) + {1 \over E_k^6} \left ( {13 \over 5040} K I - {1 \over 630}
I^3 \right ) \cdots \right ]
\label{eq:heweak}
\ee
We define a dimensionless parameter $\xi$ to characterize the field
strength
\be
\xi = {1 \over E_k} \sqrt{I \over 2}
\ee
and use the analytic expansion of this Lagrangian for small $K$
derived by Heyl and Hernquist \cite{Heyl96a}:
\be
\lag_1 = \lag_1(I,0) + K \left . \pp{\lag_1}{K} \right |_{K=0} +
\frac{K^2}{2} \left . \frac{\partial^2 \lag_1}{\partial K^2} \right
|_{K=0} + \cdots
\label{eq:lag1exp}
\ee
The first two terms of this expansion are given by
\begin{eqnarray}
\lag_1(I,0) & = & {e^2 \over h c} \frac{I}{2}
X_0\left(\frac{1}{\xi}\right), \\
 \left . \pp{\lag_1}{K} \right |_{K=0} & = & {e^2 \over h c}
\frac{1}{16 I} X_1\left(\frac{1}{\xi}\right) 
\end{eqnarray}
where
\begin{eqnarray}
X_0(x) & = & 4 \int_0^{x/2-1} \ln(\Gamma(v+1)) \d v
+ \frac{1}{3} \ln \left ( \frac{1}{x} \right )
+ 2 \ln 4\pi - (4 \ln A+\frac{5}{3} \ln 2) \nonumber \\
& & ~~ - \left [ \ln 4\pi + 1 +  \ln \left ( \frac{1}{x} \right ) \right ] x
+ \left [ \frac{3}{4} + \frac{1}{2} \ln \left ( \frac{2}{x} \right )
\right ] x^2,
\label{eq:x0anal} \\
X_1(x) & = & - 2 X_0(x) + x X_0^{(1)}(x) + \frac{2}{3} X_0^{(2)} (x) -
\frac{2}{9} \frac{1}{x^2}
\label{eq:x1anal} 
\end{eqnarray}
and
\begin{eqnarray}
X_0^{(n)}(x) &=& \frac{\d^n X_0(x)}{\d x^n}, \\
\ln A &=& \frac{1}{12} - \zeta^{(1)}(-1) \approx 0.2488.
\end{eqnarray}
where $\zeta^{(1)}(x)$ denotes the first derivative of the Riemann Zeta function.

We will treat the vacuum as a polarizable medium. 
In the Heaviside-Lorentz system, the macroscopic fields are given by
the generalized momenta conjugate to the fields \cite{Bere82}
\be
{\bf D} = \pp{\lag}{\efi} = \efi + {\bf P},~~
{\bf H} = -\pp{\lag}{\bfi} = \bfi - {\bf M},~~
{\bf P} = \pp{\lag_1}{\efi},~~
{\bf M} = \pp{\lag_1}{\bfi}.
\label{eq:genmom}
\ee
We define the vacuum dielectric and inverse magnetic permeability tensors as
follows \cite{Jack75}
\be
D_i = \epsilon_{ij} E_j,~~H_i = \mu'_{ij} B_j.
\ee
Using the definitions of $I$ and $K$, we get
\ba
\epsilon_{ij} &=& \delta_{ij} - 4 \pp{\lag_1}{I} \delta_{ij} - 32
\pp{\lag_1}{K} B_i B_j, \\
\mu'_{ij} &=& \delta_{ij} - 4 \pp{\lag_1}{I} \delta_{ij} + 32
\pp{\lag_1}{K} E_i E_j. 
\ea
If we use the weak-field limit (\eqref{heweak}), we recover the
results of Klein and Nigam \cite{Klei64a}
\ba
\epsilon_{ij} &=& \delta_{ij} + \frac{1}{45 \pi} \frac{\alpha}{B_k^2}
\left [ 2(E^2-B^2) \delta_{ij} + 7 B_i B_j \right ], \\
\mu'_{ij} &=& \delta_{ij} + \frac{1}{45 \pi} \frac{\alpha}{B_k^2}
\left [ 2(E^2-B^2) \delta_{ij} - 7 E_i E_j \right ] 
\ea
where the fine-structure constant, $\alpha=e^2/\hbar c$ in these
units.

For wrenchless ($K=0$) fields of arbitrary strength we use
\eqref{x0anal} and \eqref{x1anal} to get
\ba
\epsilon_{ij} &=& \delta_{ij} + \frac{\alpha}{2 \pi}
\left \{  \left[-2 X_0 \left(\frac{1}{\xi}\right) + \frac{1}{\xi} X_0^{(1)}
\left(\frac{1}{\xi}\right) \right ]\delta_{ij} - \frac{1}{\xi^2}
\frac{B_i B_j}{B_k^2} X_1 \left ( \frac{1}{\xi} \right ) \right \}
+{\cal O}\left[ \left (\frac{\alpha}{2\pi} \right )^2 \right],
\\
\mu'_{ij} &=& \delta_{ij} + \frac{\alpha}{2 \pi}
\left \{ \left[-2 X_0 \left(\frac{1}{\xi}\right) +\frac{1}{\xi} X_0^{(1)}
\left(\frac{1}{\xi}\right) \right ]\delta_{ij} + \frac{1}{\xi^2}
\frac{E_i E_j}{E_k^2} X_1 \left ( \frac{1}{\xi} \right ) \right \} 
+{\cal O}\left[ \left (\frac{\alpha}{2\pi} \right )^2 \right].
\ea
The expression for $\mu$ with only an external magnetic field
agrees numerically with the results of Mielnieczuk \etal \cite{Miel88}.

To examine wave propagation, we must first linearize the relations
(\eqref{genmom}) in the fields of the wave (${\tilde \efi}, {\tilde
\bfi}$)
\cite{Adle71} and obtain a second
set of matrices,
\ba
{\tilde \epsilon}_{ij} &=& \frac{\partial^2 \lag}{\partial E_i \partial E_j}, \\
& = &
\delta_{ij} - 4 \pp{\lag_1}{I} \delta_{ij}
+ 16 \frac{\partial^2 \lag_1}{\partial I^2} E_i E_j
- \left ( 64 K \frac{\partial^2 \lag_1}{\partial K^2}
+ 32 \pp{\lag_1}{K} \right ) B_i B_j 
+ 128 (\efi\cdot\bfi) \frac{\partial^2 \lag_1}{\partial I \partial
K} \left ( E_i B_j + E_j B_i \right ),
\\
{\tilde \mu}'_{ij} &=& -\frac{\partial^2 \lag}{\partial B_i \partial B_j}, \\
&=& \delta_{ij} - 4 \pp{\lag_1}{I} \delta_{ij}
- 16 \frac{\partial^2 \lag_1}{\partial I^2} B_i B_j
+ \left ( 64 K \frac{\partial^2 \lag_1}{\partial K^2}
+ 32 \pp{\lag_1}{K} \right ) E_i E_j 
+ 128 (\efi\cdot\bfi) \frac{\partial^2 \lag_1}{\partial I \partial
K} \left ( E_i B_j + E_j B_i \right ).
\ea
We use these matrices in the macroscopic Maxwell equations.  To first
order, ${\tilde \hfi} \| {\tilde \bfi}$ and ${\tilde \dfi} \| {\tilde
\efi}$, so we obtain the wave equation,
\be
\nabla^2 {\tilde \efi} - \frac{\tilde \epsilon}{{\tilde \mu}' c^2}
\frac{\partial^2 {\tilde \efi}}{\partial t^2} = 0.
\label{eq:waveE}
\ee
and similarly for ${\tilde \bfi}$. 

In \eqref{waveE}, ${\tilde \mu}'$ and ${\tilde \epsilon}$ are the
ratios of the macroscopic to the microscopic fields, \ie
${\tilde \hfi}={\tilde \mu}' {\tilde \bfi}$
The waves travel at a definite velocity
$v=c\sqrt{{\tilde \mu}'/{\tilde \epsilon}}$ and the index of
refraction is $n=\sqrt{{\tilde \epsilon}/{\tilde\mu}'}$.   

If we take an external magnetic field parallel to the $\hat{3}$ axis,
we obtain 
\ba
{\tilde \epsilon}_{ij} &=& \delta_{ij} \left \{ 1 + \frac{\alpha}{2 \pi}
\left [ -2 X_0 \left(\frac{1}{\xi}\right) + \frac{1}{\xi} X_0^{(1)}
\left(\frac{1}{\xi}\right) \right ] \right \} - \delta_{i3}
\delta_{j3}\frac{\alpha}{2 \pi}  X_1 \left ( \frac{1}{\xi} \right )
+{\cal O}\left[ \left (\frac{\alpha}{2\pi} \right )^2 \right],
\\
{\tilde \mu}'_{ij} &=& \delta_{ij} \left \{ 1 + \frac{\alpha}{2 \pi}
\left [ -2 X_0 \left(\frac{1}{\xi}\right) + \frac{1}{\xi} X_0^{(1)}
\left(\frac{1}{\xi}\right) \right ] \right \}
- \delta_{i3} \delta_{j3} \frac{\alpha}{2 \pi} 
\left [  X_0^{(2)} \left ( \frac{1}{\xi} \right ) \xi^{-2} -
  X_0^{(1)} \left ( \frac{1}{\xi} \right ) \xi^{-1} \right ]
+{\cal O}\left[ \left (\frac{\alpha}{2\pi} \right )^2 \right].
\ea
In this case, we have the magnetic field of the wave either
perpendicular to the plane containing the external magnetic field
and the direction of propagation
(${\bf k}$), $\perp$ mode, or in that 
plane, $\|$ mode \cite{Bere82}.  For the $\perp$ mode, we obtain
\be
n_\perp = 1 - \frac{\alpha}{4\pi} X_1 \left(\frac{1}{\xi}\right)
\sin^2 \theta + {\cal O}\left[ \left (\frac{\alpha}{2\pi} \right )^2
\right]
\label{eq:nperpb}
\ee
where $\theta$ is the angle between the direction of propagation and
the external field.  And for the $\|$ mode, we obtain
\be
n_\| = 1 + \frac{\alpha}{4 \pi}
\left [ X_0^{(2)} \left ( \frac{1}{\xi} \right ) \xi^{-2} -
  X_0^{(1)} \left ( \frac{1}{\xi} \right ) \xi^{-1} \right ] \sin^2 \theta
+ {\cal O}\left[ \left (\frac{\alpha}{2\pi} \right )^2 \right ].
\label{eq:nparab}
\ee
The expressions for $n_\|,n_\perp$ obtained here are equivalent to
those obtained by Tsai and Erber\cite{Tsai75} through direct
calculation of the photon propagator to one-loop accuracy in the
presence of the external field.

If we take an external electric field parallel to the $\hat{3}$ axis,
we obtain 
\ba
{\tilde \epsilon}_{ij} &=& \delta_{ij} \left \{ 1 + \frac{\alpha}{2 \pi}
\left [ -2 X_0 \left(\frac{1}{\xi}\right) + \frac{1}{\xi} X_0^{(1)}
\left(\frac{1}{\xi}\right) \right ] \right \}
- \delta_{i3} \delta_{j3} \frac{\alpha}{2 \pi} 
\left [  X_0^{(2)} \left ( \frac{1}{\xi} \right ) \xi^{-2} -
  X_0^{(1)} \left ( \frac{1}{\xi} \right ) \xi^{-1} \right ]
+{\cal O}\left[ \left (\frac{\alpha}{2\pi} \right )^2 \right],
\\
{\tilde \mu}'_{ij} &=& \delta_{ij} \left \{ 1 + \frac{\alpha}{2 \pi}
\left [ -2 X_0 \left(\frac{1}{\xi}\right) + \frac{1}{\xi} X_0^{(1)}
\left(\frac{1}{\xi}\right) \right ] \right \} - \delta_{i3}
\delta_{j3} \frac{\alpha}{2 \pi} X_1 \left ( \frac{1}{\xi} \right )
+{\cal O}\left[ \left (\frac{\alpha}{2\pi} \right )^2 \right].
\ea
In this case, the propagation modes have the electric field (${\tilde \efi})$
either in the ${\bf k}-\efi$ plane ($\|$ mode) or 
or perpendicular to the plane.  For an external electric field, we define
\be
\xi = i y = i \frac{E}{E_k} 
\ee
and substitute this into \eqref{nperpb} and \eqref{nparab}.  This
yields indices of refraction
\ba
n_\perp &=& 1 + \frac{\alpha}{4\pi} X_1 \left(\frac{1}{i y}\right)
\sin^2 \theta + {\cal O}\left[ \left (\frac{\alpha}{2\pi} \right )^2
\right],
\label{eq:nperpe}
\\
n_\| &=& 1 + \frac{\alpha}{4 \pi}
\left [ X_0^{(2)} \left ( -\frac{i}{y} \right ) y^{-2} -
  i X_0^{(1)} \left ( -\frac{i}{y} \right ) y^{-1} \right ] \sin^2 \theta
+ {\cal O}\left[ \left (\frac{\alpha}{2\pi} \right )^2 \right ]
\label{eq:nparae}
\ea
where $\theta$ again refers to the angle between the direction of
propagation and the external electric field.

In the weak-field limit, we have \cite{Heyl96a}
\ba
X_1 \left ( \frac{1}{\xi} \right ) &=& -\frac{14}{45} \xi^2 + {\cal O}
(\xi^4), \\
 X_0^{(2)} \left ( \frac{1}{\xi} \right ) \xi^{-2} -
  X_0^{(1)} \left ( \frac{1}{\xi} \right ) \xi^{-1} &=& \frac{8}{45} \xi^2
 + {\cal O}(\xi^4).
\ea
An external electric field gives $\xi^2<0$ and an external magnetic
field gives $\xi^2>0$, therefore $n_\perp,n_\|>1$ in the weak-field limit for
both cases.  Using this limit in \eqref{nparab} and
\eqref{nperpb} yields weak-field expressions for the index of refraction
in a magnetic field in agreement with earlier work \cite{Adle71,Bere82}.

\subsection{Series and asymptotic expressions}

To calculate the indices of refraction in the weak and strong field
limit, we use the expansions of Heyl \& Hernquist\cite{Heyl96a}.
For an external magnetic field, in the weak-field limit ($\xi < 0.5$),
\ba
n_\perp &=& 1 + \frac{\alpha}{4 \pi} \sin^2\theta \left [ \frac{14}{45} \xi^2
- \frac{1}{3} \sum_{j=2}^\infty \frac{2^{2j} \left (6 B_{2(j+1)} -
(2j+1) B_{2j} \right )}{j(2j+1)} \xi^{2j} \right ]
+ {\cal O}\left[ \left (\frac{\alpha}{2\pi} \right )^2 \right], \\
n_\| &=& 1 - \frac{\alpha}{4\pi} \sin^2\theta \sum_{j=1}^\infty 
\frac{2^{2(j+1)} B_{2(j+1)}}{2j+1} \xi^{2j}  
+ {\cal O}\left[ \left (\frac{\alpha}{2\pi} \right )^2 \right].
\ea
In the strong-field limit ($\xi > 0.5$), we obtain
\ba
n_\perp &=& 1 + \frac{\alpha}{4\pi} \sin^2\theta \Biggr [ \frac{2}{3} \xi
- \left ( 8 \ln A - \frac{1}{3} - \frac{2}{3} \gamma \right ) 
- \left ( \ln\pi + \frac{1}{18} \pi^2 - 2 - \ln \xi \right ) \xi^{-1}
- \left ( -\frac{1}{2} - \frac{1}{6} \zeta(3) \right ) \xi^{-2} 
\nonumber \\*
& & ~~~ - \sum_{j=3}^\infty \frac{(-1)^{j-1}}{2^{j-2}} 
 \left ( \frac{j-2}{j(j-1)} \zeta(j-1) + \frac{1}{6}
\zeta(j+1) \right ) \xi^{-j} \Biggr ] 
+ {\cal O}\left[ \left (\frac{\alpha}{2\pi} \right )^2 \right ], \\
n_\| &=& 1 + \frac{\alpha}{4\pi} \sin^2\theta \Biggr [ 
\frac{2}{3} - \frac{\ln\xi + 1 -\ln\pi}{\xi} - \frac{1}{\xi^2}
+ \sum_{j=3}^\infty \frac{(-1)^{j-1}}{2^{j-2}} \frac{j-2}{j-1} 
\zeta(j-1) \xi^{-j}
\Biggr ]
+ {\cal O}\left[ \left (\frac{\alpha}{2\pi} \right )^2 \right ],
\ea
where $\gamma$ is Euler's constant.

For an external electric field, in the weak-field
limit ($y < 0.5$) we obtain,
\ba
n_\perp &=& 1 + \frac{\alpha}{4\pi} \left [ \frac{14}{45} y^2
+ \frac{1}{3} \sum_{j=2}^\infty \frac{(-1)^{j} 2^{2j} \left (6 B_{2(j+1)} -
(2j+1) B_{2j} \right )}{j(2j+1)} y^{2j} \right ]
+ {\cal O}\left[ \left (\frac{\alpha}{2\pi} \right )^2 \right], \\
n_\| &=& 1 + \frac{\alpha}{4\pi} \sin^2\theta \sum_{j=1}^\infty 
\frac{(-1)^j 2^{2(j+1)} B_{2(j+1)}}{2j+1} y^{2j}  
+ {\cal O}\left[ \left (\frac{\alpha}{2\pi} \right )^2 \right], \\
\ea
and in the strong-field limit ($y > 0.5$)
\ba
n_\perp &=& 1 + \frac{\alpha}{4\pi} \sin^2\theta \Biggr [ - i \frac{2}{3} y
+ \left ( 8 \ln A - \frac{1}{3} - \frac{2}{3} \gamma \right ) 
- i \left ( \ln\pi + \frac{1}{18} \pi^2 - 2 - \ln (i y) \right ) y^{-1}
- \left ( -\frac{1}{2} - \frac{1}{6} \zeta(3) \right ) y^{-2} 
\nonumber \\*
& & ~~~ + \sum_{j=3}^\infty \frac{(-1)^{j-1}}{2^{j-2}} 
 \left ( \frac{j-2}{j(j-1)} \zeta(j-1) + \frac{1}{6}
\zeta(j+1) \right ) (i y)^{-j} \Biggr ] 
+ {\cal O}\left[ \left (\frac{\alpha}{2\pi} \right )^2 \right ],
\\
n_\| &=& 1 - \frac{\alpha}{4\pi} \sin^2\theta \Biggr [ 
\frac{2}{3} + i\frac{\ln(iy) + 1 -\ln\pi}{y} + \frac{1}{y^2}
+ \sum_{j=3}^\infty \frac{(-1)^{j-1}}{2^{j-2}} \frac{j-2}{j-1} 
\zeta(j-1) (i y)^{-j}
\Biggr ]
+ {\cal O}\left[ \left (\frac{\alpha}{2\pi} \right )^2 \right ].
\ea
From this equation, it is apparent that the index of refraction
acquires an imaginary part in strong electric fields.

\section{Birefrigence}

\begin{figure}
\plotone{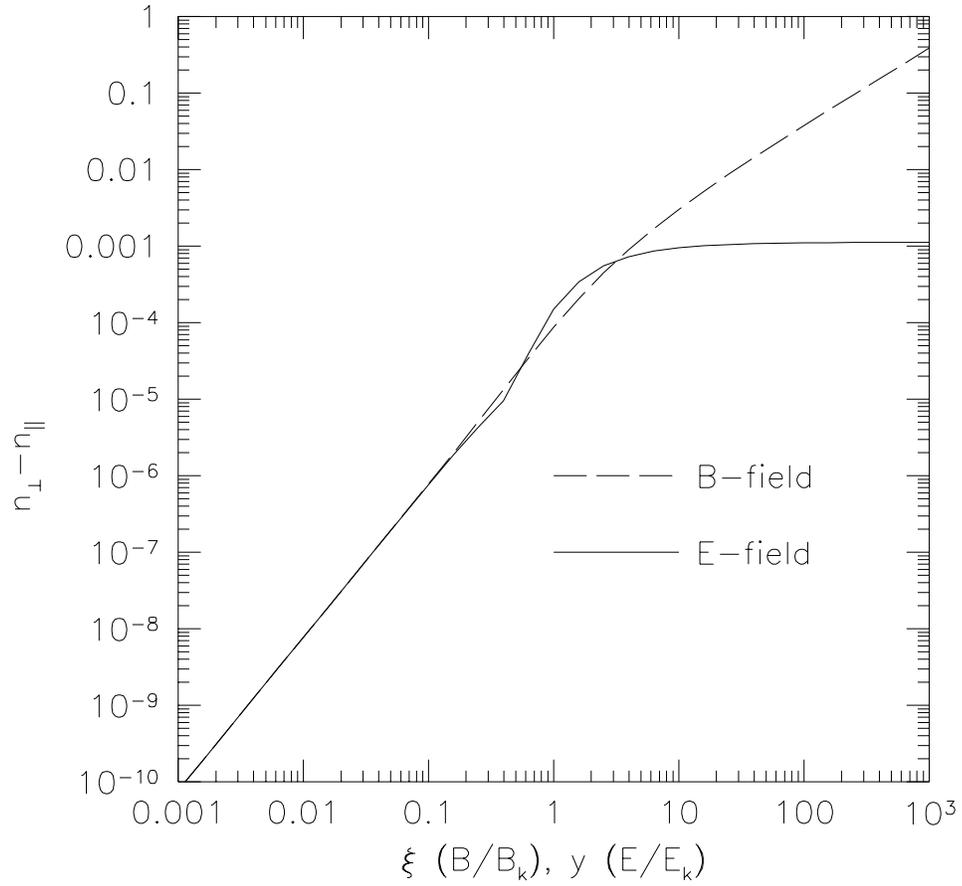} \nobreak
\caption{The difference between the index of refraction of the
parallel and perpendicular polarizations for light
travelling through external electric or magnetic fields.}
\label{fig:biref}
\end{figure}

In general, the birefringence is quantified by the difference of the
indexes of refraction for the two modes of propagation,
\be
n_\perp - n_\| = \pm \frac{\alpha}{4\pi} \left [
X_0^{(1)} \left ( \frac{1}{\xi} \right ) \xi^{-1}
- X_0^{(2)} \left ( \frac{1}{\xi} \right ) \xi^{-2}
- X_1 \left ( \frac{1}{\xi}\right ) \right ] \sin^2\theta
+ {\cal O}\left[ \left (\frac{\alpha}{2\pi} \right )^2 \right]
\ee
where the upper sign is for the magnetized case and the lower for
the electrified case.  \figref{biref} depicts the indices of
refraction for these two cases.  

\section{Dichroism}

The analytic properties of the function $n_\|(\xi)$ can be used to
estimate the dichroic properties of a magnetized or electrified
vacuum.  In a external electric field we have $\xi=i E/E_k=i y$, while in
a magnetic field $\xi=B/B_k$.  $n_\|(\xi)$ is real for real arguments;
however for imaginary $\xi$, $n_\|(\xi)$ acquires an imaginary
part.  Classically, this imaginary part may be related to the
attentuation length of a plane wave traversing the vacuum
\be
l = \frac{2 \pi \lambda}{\IM n}
\ee
where $\lambda$ is the wavelength of the radiation.  In quantum field
theory, the imaginary part of $n$ is related to the imaginary
part of the photon polarization operator and therefore the
cross-section for one-photon pair production.

\begin{figure}
\plotone{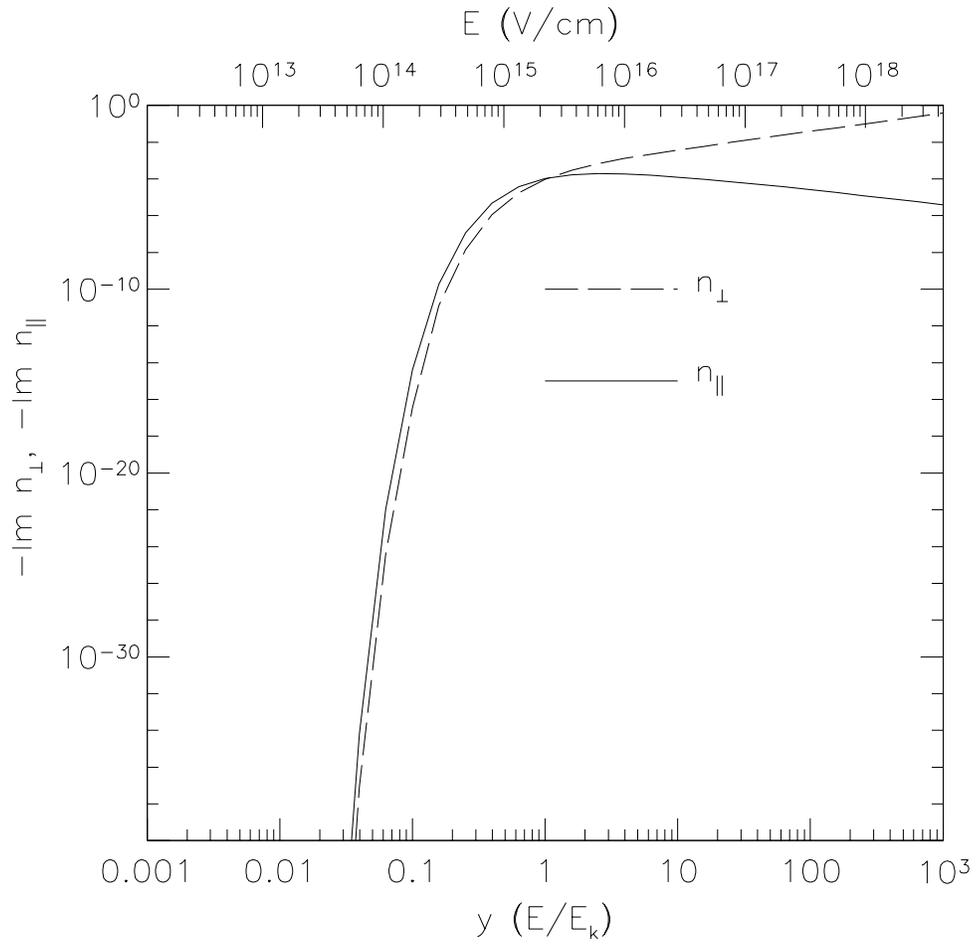} \nobreak
\caption{The imaginary part of the index of refraction for
perpendicular and parallel propagation modes for light travelling
through an external electric field.}
\label{fig:imne}
\end{figure}

In general the imaginary part for the two polarization modes is
\ba
\IM n_\perp &=& \frac{\alpha}{4 \pi} \sin^2\theta \IM X_1 \left ( -\frac{i}{y} \right )
+ {\cal O}\left[\left(\frac{\alpha}{2\pi}\right)^2 \right], \\
\IM n_\| &=& \frac{\alpha}{4 \pi} \sin^2\theta \left [
\IM X_0^{(2)} \left ( -\frac{i}{y} \right ) y^{-2} -
\RE X_0^{(1)} \left ( -\frac{i}{y} \right ) y^{-1} \right ]
+ {\cal O}\left[\left(\frac{\alpha}{2\pi}\right)^2 \right]. 
\ea
These are conveniently calculated by evaluating the imaginary part
of $X_0(x)$ for imaginary values of $x$ by integrating
around the poles of \eqref{lag1def} \cite{Heyl96a,Itzy80},
\be
\IM X_0(x) = - \frac{1}{\pi} \sum_{n=1}^\infty \frac{e^{-i \pi n
x}}{n^2} =
- \frac{1}{\pi} e^{-\pi/y}~_{1,1,1}F_{2,2}\left (e^{-\pi/y}\right) 
\ee
where $F$ is a generalized hypergeometric function. 
Using \eqref{x1anal} to calculate $\IM X_1(x)$, yielding for the
indices of refraction,
\ba
\IM n_\perp &=& \frac{\alpha}{4 \pi} \sin^2\theta \sum_{n=1}^\infty
\left ( \frac{2}{3}\pi + \frac{1}{n} \frac{1}{y} + \frac{1}{n^2} \frac{2}{\pi} 
 \right ) e^{-n \pi/y}
+ {\cal O}\left[\left(\frac{\alpha}{2\pi}\right)^2 \right], \\
&=& \frac{\alpha}{4\pi} \sin^2\theta \left  [
\frac{2}{3}\pi \left( e^{\pi/y} - 1 \right )^{-1}
- \frac{1}{y} \ln \left ( 1 - e^{-\pi/y}\right ) 
+ \frac{2}{\pi} e^{-\pi/y}~_{1,1,1}F_{2,2}\left (e^{-\pi/y}\right)
\right ]
+ {\cal O}\left[\left(\frac{\alpha}{2\pi}\right)^2 \right],
\label{eq:imnperp} \\
\IM n_\| &=& \frac{\alpha}{4 \pi} \sin^2\theta \sum_{n=1}^\infty
\left ( \frac{\pi}{y^2} + \frac{1}{n} \frac{1}{y} \right ) e^{-n \pi/y}
+ {\cal O}\left[\left(\frac{\alpha}{2\pi}\right)^2 \right], \\
&=& \frac{\alpha}{4\pi} \sin^2\theta \left  [
\frac{\pi}{y^2} \left( e^{\pi/y} - 1 \right )^{-1}
- \frac{1}{y} \ln \left ( 1 - e^{-\pi/y}\right )
\right ]
+ {\cal O}\left[\left(\frac{\alpha}{2\pi}\right)^2 \right]
\label{eq:imnpara}
\ea
\figref{imne} depicts the imaginary part of the index of
refraction as a function of field strength.

In the weak-field limit, the imaginary part of the index of
refraction is exponentially small as Klein and Nigam \cite{Klei64b} found.
However, our result is larger by a factor of $1/y$ in this limit and
is more complicated.  The error occurs between their Eq. 5 and
Eq. 6.  First, they have neglected the real part of the integral,
and as in Ref.\cite{Klei64a}, they have calculated $\mu'_{ij}$
and used it as $\mu_{ij}$. These errors are not important for this
application.  However, their function $\Phi_2(x)$ has not been calculated
correctly.  By examination of their Eq.6, we see that
\be
\pp{\lag}{K} = - i \frac{\alpha}{2} \frac{1}{16 I} \Phi_2(x)
\ee
so
\be
\Phi_2(x) = -\frac{1}{\pi} \IM X_1 \left ( \frac{\pi}{i x} \right )
\ee
which from examination of \eqref{imnperp} is significantly more
complicated than their expression.

\section{Conclusions}

Using a closed form expression for the Heisenberg-Euler effective
Lagrangian for quantum electrodynamics in wrenchless ($K=0$) fields,
we have calculated general expressions for the index of refraction of
a slowly-varying electromagnetic field, and evaluated these
expressions for the simple cases of a pure electric or magnetic field.
Our results agree with some previous work \cite{Adle71,Bere82,Miel88,Tsai75}
in the appropriate limits.  We expect these results to be of general
utility especially in the study of light propagation in the vicinity
of strongly magnetized neutron stars.

\acknowledgements

This material is based upon work supported under a National Science
Foundation Graduate Fellowship. L.H. thanks the National Science
Foundation for support under the Presidential Faculty Fellows Program.

\end{document}